%% file: manuscript.tex
\documentclass[sigconf, screen]{acmart}

\AtBeginDocument{%
  \providecommand\BibTeX{{%
    \normalfont B\kern-0.5em{\scshape i\kern-0.25em b}\kern-0.8em\TeX}}}

\setcopyright{acmcopyright}

\copyrightyear{2026}
\acmYear{2026}
\setcopyright{cc}
\setcctype{by}
\acmConference[CHI '26 Workshop]{Bias4Trust: Understanding, Mitigating, and Leveraging Cognitive Biases to Calibrate Trust in Evolving AI Systems at CHI '26}{April 13--17, 2026}{Barcelona, Spain}
\acmBooktitle{Bias4Trust Workshop on Understanding, Mitigating, and Leveraging Cognitive Biases to Calibrate Trust in Evolving AI Systems at the 2026 CHI Conference on Human Factors in Computing Systems (CHI '26), April 13--17, 2026, Barcelona, Spain}
\acmPrice{}
\acmDOI{XX.XXXX/XXXXXXX.XXXXXXX}
\acmISBN{XXX-X-XXXX-XXXX-X/2026/04}

\begin{document}

\title{Multi-Agent Consensus as a Cognitive Bias Trigger in Human-AI Interaction
}

\author{Soohwan Lee}
\orcid{0000-0001-8652-3408}
\affiliation{%
  \institution{Department of Design, UNIST}
  \city{Ulsan}
  \country{Republic of Korea}}
\email{soohwanlee@unist.ac.kr}

\author{Kyungho Lee}
\orcid{0000-0002-1292-3422}
\affiliation{\institution{Department of Design, UNIST}
\city{Ulsan}
\country{Republic of Korea}}
\email{kyungho@unist.ac.kr}

\begin{abstract}
As multi-agent AI systems become more common, users increasingly encounter not a single AI voice but a collective one. This shift introduces social dynamics, such as consensus, dissent, and gradual convergence, that can trigger cognitive biases and distort human judgment. We present findings from a controlled experiment (N = 127) comparing three multi-agent configurations: Majority, Minority, and Diffusion. Quantitative results show that majority consensus accelerates opinion change and inflates confidence, consistent with social proof and bandwagon heuristics. Minority dissent slows this process and promotes more deliberative engagement. Qualitative analysis identifies three interpretive trajectories: reinforcing, aligning, and oscillating, shaped by how users interpret agent independence and group dynamics over time. These findings suggest that agent agreement structure, independent of content, functions as a bias-relevant signal in LLM interactions. We hope this work contributes to the Bias4Trust agenda by grounding multi-agent social influence as a concrete and designable source of bias in human-AI interaction.

\end{abstract}



\maketitle

\input{sections/01_Introduction}

\begin{acks}
This research was partially supported by a grant from the Korea Institute for Advancement of Technology (KIAT) funded by the Government of Korea (MOTIE) (P0025495, Establishment of Infrastructure for Integrated Utilization of Design Industry Data).
\end{acks}

\bibliographystyle{ACM-Reference-Format}

\bibliography{SIMA}

\input{sections/99_appendix}

\end{document}

%% file: sections/01_Introduction.tex
\section{Introduction}
Human-AI interaction research has predominantly operated under the "Computers Are Social Actors" (CASA) paradigm, focusing on dyads between a single user and a single agent \cite{reevesMediaEquationHow1996,nassComputersAreSocial1994,capelWhatHumanCenteredHumanCentered2023,laiScienceHumanAIDecision2023}. However, the rapid proliferation of Large Language Models (LLMs) is shifting this landscape toward artificial crowds, where users interact with multiple agents simultaneously in collaborative and social settings \cite{parkGenerativeAgentsInteractive2023a,seeberMachinesTeammatesResearch2020,fourneyMagenticOneGeneralistMultiAgent2024, dibiaAutoGenStudioNoCode2024}. This transition is critical because multi-agent systems possess the unique potential to form \textit{group dynamics}—such as consensus, dissent, and coalition—that can significantly amplify the impact of cognitive biases on human decision-making.

Unlike single-agent advice, multi-agent collectives generate powerful social signals that trigger deep-seated heuristics. For instance, a "synthetic consensus" among agents can exploit the \textit{bandwagon effect} or \textit{consensus heuristic}, leading users to uncritically conform to machine-generated views under the false assumption of independent validation \cite{songGreaterSumIts2025,songMultiAgentsAreSocial2025}. Conversely, consistent dissent from a minority agent can disrupt automatic processing and trigger reflective (System 2) thinking, a phenomenon known as minority influence \cite{moscoviciStudiesSocialInfluence1976,moscoviciTheoryConversionBehavior1980}. While temporal dynamics of minority influence have been examined in social psychology \cite{prislinSocialChangeAftermath2005}, how users interpret these signals over time in LLM-based multi-agent settings remains comparatively underexplored.

In this work, we investigate how multi-agent configurations shape user trust and reliance through the lens of cognitive bias. Drawing on a controlled experiment ($N=127$) with Majority, Minority, and Diffusion conditions, we analyze both quantitative behavioral shifts and qualitative interpretive trajectories \cite{leeComplianceAndConversion2026}. Our findings reveal that while the majority consensus accelerates the largest opinion change, minority dissent and diffusion can induce a conversion-like mechanism, depending on how users attribute the agents' intent.

At the Bias4Trust workshop, we aim to provoke a discussion on "Understanding and Mapping Biases in Human–AI Interaction"—how we can leverage multi-agent friction to mitigate over-reliance on synthetic consensus and support more calibrated trust in AI systems.

\section{Brief Methodology}
We conducted a split-plot mixed experiment with 127 Prolific participants (UK/US), randomly assigned to one of three between-subject conditions (Majority, Minority, or Diffusion) and completing both a normative and an informative task in a counterbalanced order (about 30 minutes; IRB-exempt: UNISTIRB-25-062-C).
Each task began with participants reporting their initial opinion and confidence (T0), followed by four sequential interaction cycles with three GPT-4o agents via a group-chat interface (\autoref{fig:systemInterface}). Before and after each chat cycle (C1-C4), participants re-reported their opinion (-50 to +50) and confidence (0-100) to track change over time (\autoref{fig:experimentalProcedure}).

Manipulating conditions affected how agent alignment evolved across cycles. In the Majority condition, all three agents consistently opposed the participant throughout. In the Minority condition, one agent consistently dissented while the other two supported the participant. In the Diffusion condition, the session began like Minority, but one supportive agent switched to opposition at Cycle 3 and the other at Cycle 4, creating a gradual transition from minority dissent to majority opposition (\autoref{fig:experimentalDesign}).

Normative tasks involved value-based judgments (e.g., policy preferences) aimed at social-normative influence; informative tasks involved fact-based statements aimed at informational influence. Measures included repeated opinion and confidence change (directional and absolute), stance reversals, perceived compliance, perceived conversion, and agent credibility evaluations, supplemented by open-ended responses on participants' reasoning and perceptions of social influence. Full methodological details are reported in \cite{leeComplianceAndConversion2026}

\section{Key Findings}
\subsection{Quantitative Findings: Consensus Acts as a Fast Trust Cue in Informative Tasks}
We analyzed repeated measures of opinion and confidence across time using linear mixed-effects models. We included condition (Majority vs. Minority vs. Diffusion), task type (informative vs. normative), and time (T0--T4), plus their interactions, with random intercepts for participants. We followed up key interactions with estimated marginal means and Bonferroni post-hoc tests. We also tested stance reversals (opinion sign flips) using chi-square tests. We report standardized effect sizes (Hedges' $g$) to convey practical magnitude. Details are shown in \autoref{tab:lmm_opinion_confidence_full}

Across measures, task type shaped the pattern. Informative tasks showed a clear separation between conditions, while normative tasks showed smaller, less consistent movements. This difference suggests that participants relied more on social cues when judging truth-like claims than when expressing preferences.

In informative tasks, the Majority condition produced the fastest and largest opinion updating. Participants shifted earlier in the sequence and showed larger absolute opinion change than in Diffusion (medium-to-large effects, $g \approx 0.68$--$1.06$ across time points). Minority produced the smallest and slowest opinion change. It often delays updating rather than driving it, which aligns with the idea that a single dissenting voice can reduce immediate conformity pressure.

Confidence showed a related but not identical pattern. Majority increased confidence quickly and strongly at midpoints (Majority $>$ Minority at T2--T3, $g \approx 0.67$--$0.80$). Diffusion showed a later and more sustained confidence rise, and it exceeded Minority by the end of the sequence (T4, $g \approx 0.77$). This result suggests that gradual convergence can feel like accumulating evidence, even when the underlying information remains unchanged.

Multi-agent agreement can function as a high-salience heuristic in LLM interactions, especially for informative tasks. Majority consensus can compress deliberation and accelerate belief updating, while diffusion can inflate confidence through a ``growing agreement'' narrative. Designers should treat the agreement structure as a trust-relevant signal, make it legible, and add friction to prompt independent checking before users lock in beliefs and confidence.

\subsection{Qualitative Findings: Interpretive Trajectories Beyond Persuasion}

We conducted a thematic analysis of participants' post-task open-ended reflections \cite{braunUsingThematicAnalysis2006}. Participants did not consistently move toward the agents. Instead, they showed three recurring trajectories over time: reinforcing, aligning, and oscillating. Reinforcing often reflected active engagement, where participants checked claims and then strengthened their initial stance. Aligning often followed credible information updates, especially when participants started from uncertainty. Oscillating captured repeated recalibration, where participants alternated between accepting plausible content and questioning the interaction process (\autoref{tab:trajectories}).

Participants explained these trajectories through four interpretive lenses: interactional legitimacy (whether the group listens and responds), collective reasoning strength (whether arguments stay coherent and checkable), independent plurality (whether voices feel independent rather than a coalition), and interpretation of group dynamics (what convergence or switching means). When agents sounded repetitive, unresponsive, or scripted, participants often inferred coordinated pressure and reported reactance, which reinforced their stance. In minority-like settings, some participants discounted supportive agents as ``yes-men'' and treated a dissenting voice as more independent, so they shifted when dissent offered more verifiable reasoning. In diffusion, convergence increased confidence for some participants, but unexplained switching also raised suspicion about bandwagoning or orchestration, which destabilized trust and prompted renewed resistance.

Trust miscalibration in multi-agent LLMs can take multiple forms, including fast alignment, reactance-driven reinforcement, and oscillation driven by process ambiguity. Further discussions can map these outcomes as distinct bias pathways and explore interface supports that make group processes legible, for example, cues for independence and short change records that state why an agent changed its stance.

\section{Discussion}

Our results suggest that multi-agent social signals can shape judgment not only through content but also through perceived group dynamics over time. In the majority condition, participants tended to reach a decision and stabilize confidence earlier. This pattern aligns with biases that treat consensus as a cue for correctness, such as social proof and the bandwagon effect. In an LLM interface, consensus also reduces effort. Users can rely on a simple heuristic, “many agents agree, so this seems right,” which fits the Need to Act Fast cluster. At the same time, the reinforcing trajectory points to a more complex interaction with confirmation bias and motivated reasoning. Even when agents present counter-attitudinal claims, users may treat them as a threat and respond by strengthening prior beliefs. This mechanism resembles reactance and belief-defense processes rather than a classic filter bubble that relies on receiving only agreeable information.

The minority and diffusion conditions exhibit distinct cognitive profiles. They slowed down opinion change and delayed confidence stabilization, which may reflect increased deliberation and reduced reliance on shortcuts to consensus. This dynamic connects to the Not Enough Meaning cluster. When agents disagree or shift, users face ambiguity and must interpret “what this disagreement means.” They may engage more in sensemaking, but they also face higher cognitive effort. The diffusion trajectory adds another layer by treating social change as a signal. A gradual shift can create an implicit narrative of persuasion and momentum, inviting anchoring to early positions and later adjustment based on perceived movement. These patterns indicate that trust calibration in multi-agent settings may depend as much on temporal social structure as on factual accuracy.

These dynamics open risky scenarios. First, a system can stage “consensus theater” to steer users in political or commercial contexts. Our findings suggest that coordinated agents can create artificial majority pressure, pushing rapid compliance even without strong evidence. Second, an attacker can exploit reinforcing dynamics as a new dark pattern. This raises the concern that by injecting obviously opposing or extreme “counter-evidence,” the system can trigger reactance and strengthen a target belief. This tactic can amplify polarization while appearing balanced. Third, minority and diffusion settings can support reflection when a truthful minority challenges errors, but they also create a pathway for misinformation. A small number of agents can introduce a plausible but false source, and other agents can cite, echo, or converge on it over time \cite{choiEmpiricalStudyGroup2025}. This can resemble a form of trust laundering, where users trust the claim because multiple agents repeat it, not because the evidence holds. In domains such as health, finance, or civic decision-making, this failure mode can lead to high-impact miscalibration.

We propose design provocations that treat multi-agent influence as an interface problem. 1) Independence and provenance cues: show whether agents share tools, memory, or retrieval sources, and highlight when agents recycle the same evidence. 2) Cross-agent firewalls: prevent agents from updating beliefs based only on other agents, and require external evidence for shifts. 3) Adaptive cognitive friction: when the system detects rapid convergence or rising confidence, it can add a brief reflection step, request a justification, or delay the reveal of agreement signals. Importantly, friction raises cognitive workload. Our minority condition suggests that deliberation can slow decisions, so designers should tune friction to context. High-stakes tasks may warrant a greater workload, while low-stakes tasks may require lighter scaffolding \cite{leeConversationalAgentsCatalysts2025b}. A flexible balance between critical thinking and cognitive effort can help users calibrate trust without making the interaction unusable.

\section{Conclusion}
Multi-agent AI systems introduce social dynamics that single-agent models do not capture. Our study shows that agreement structure, not just content, shapes how users update beliefs and calibrate trust. Majority consensus can compress deliberation and drive rapid opinion change, while minority dissent and gradual diffusion can slow this process and invite more reflective engagement. These dynamics map onto known cognitive bias clusters, including social proof, the bandwagon effect, and anchoring, and they carry real risks such as consensus theater and trust laundering. As multi-agent interfaces become more common, designers need to treat agent agreement patterns as trust-relevant signals. Practical responses include independence cues, provenance transparency, and adaptive friction. We hope this work contributes to a shared research agenda for bias-aware design in multi-agent AI systems.

%% file: sections/99_appendix.tex
\appendix
\onecolumn
\clearpage
\section{Details of Quantitative Analysis}
\begin{table*}[htbp]
\centering
\caption{Linear mixed-effects models of signed and absolute changes in opinion and confidence relative to $T_0$ ($\Delta$Opinion, $|\Delta$Opinion$|$, $\Delta$Confidence, $|\Delta$Confidence$|$). Fixed effects are experimental condition (Majority, Minority, Diffusion), task type (Normative vs.\ Informative), time step ($T_1$–$T_4$), their interactions, and standardized covariates (Susceptibility to Interpersonal Influence(SII), Need for Cognition(NFC), AI acceptance). All categorical predictors (Condition, Task, Time) are effect-coded using sum-to-zero contrasts (\texttt{contr.sum}). The intercept corresponds to the grand mean across all levels, and each level coefficient (e.g., \emph{Condition: Majority}, \emph{Time: $T_1$}) indicates a deviation from this grand mean rather than a difference from a single reference category. For the binary Task factor, Normative is coded $+1$ and Informative $-1$, so the coefficient is half the difference between the two task types. For the three-level Condition factor (Majority, Minority, Diffusion), two coefficients are shown (Majority, Minority); each is that condition’s deviation from the grand mean, and the effect for Diffusion is given implicitly as the negative sum of the other two. For Time, $T_1$–$T_3$ are shown explicitly and the effect for $T_4$ is implied by the sum-to-zero constraint. Models were fit by maximum likelihood with random intercepts for participants. Coefficients are unstandardized estimates with standard errors in parentheses. Stars denote significance (* $p<.05$, ** $p<.01$, *** $p<.001$).}
\Description{The table summarizes results from four linear mixed-effects models examining signed and absolute changes in opinion and confidence relative to baseline. Each column presents unstandardized coefficients, standard errors, and p-values for one of the four dependent variables. Predictors include condition (Majority, Minority), task type, time steps, their interactions, and standardized covariates such as social influence susceptibility, need for cognition, and AI acceptance. Across outcomes, the intercept and task type show consistent significant effects, and time effects are typically small except at later steps. Several interaction terms involving time and condition also reach significance, especially for absolute opinion and confidence changes, indicating increasing divergence across conditions as time progresses. Model summaries at the bottom report the number of observations, participant counts, random-effect variances, and AIC/BIC values.}
\scriptsize
\setlength{\tabcolsep}{3.8pt}
\renewcommand{\arraystretch}{0.95}
\begin{tabular}{lcccccccc}
\toprule
 & \multicolumn{2}{c}{$\Delta$Opinion (signed)} & \multicolumn{2}{c}{$|\Delta$Opinion$|$} & \multicolumn{2}{c}{$\Delta$Confidence (signed)} & \multicolumn{2}{c}{$|\Delta$Confidence$|$} \\
\cmidrule(lr){2-3}\cmidrule(lr){4-5}\cmidrule(lr){6-7}\cmidrule(lr){8-9}
Predictor & $\beta$ (SE) & $p$ & $\beta$ (SE) & $p$ & $\beta$ (SE) & $p$ & $\beta$ (SE) & $p$ \\
\midrule
Intercept & -4.21 (1.43) & .004** & 17.50 (1.00) & $<.001^{***}$ & 11.36 (1.31) & $<.001^{***}$ & 17.48 (1.02) & $<.001^{***}$ \\
Condition: Majority & -4.03 (2.06) & .052 & 3.68 (1.44) & .012* & -0.79 (1.88) & .677 & 2.97 (1.46) & .044* \\
Condition: Minority & -1.11 (2.05) & .589 & -0.26 (1.43) & .858 & -2.10 (1.88) & .266 & -2.33 (1.46) & .114 \\
Task: Normative & 5.16 (0.64) & $<.001^{***}$ & -6.70 (0.46) & $<.001^{***}$ & -4.07 (0.56) & $<.001^{***}$ & -6.22 (0.45) & $<.001^{***}$ \\
Time: $T_1$ & 0.70 (1.11) & .530 & -5.60 (0.80) & $<.001^{***}$ & -5.21 (0.97) & $<.001^{***}$ & -5.61 (0.78) & $<.001^{***}$ \\
Time: $T_2$ & 1.34 (1.11) & .228 & -1.93 (0.80) & .016* & -1.78 (0.97) & .067 & -1.34 (0.78) & .086 \\
Time: $T_3$ & 0.28 (1.11) & .799 & 1.91 (0.80) & .017* & 2.32 (0.97) & .017* & 1.72 (0.78) & .028* \\
SII (z) & -0.13 (1.46) & .927 & -0.29 (1.02) & .780 & 0.45 (1.34) & .738 & 0.46 (1.04) & .663 \\
NFC (z) & -1.71 (1.46) & .243 & -0.06 (1.02) & .950 & -3.68 (1.33) & .007** & -2.22 (1.04) & .034* \\
AI acceptance (z) & -1.95 (1.48) & .189 & 0.77 (1.03) & .454 & 0.18 (1.35) & .896 & 1.19 (1.05) & .260 \\
Condition: Majority $\times$ Task: Normative & -1.27 (0.92) & .166 & -3.37 (0.66) & $<.001^{***}$ & -1.71 (0.80) & .033* & -1.23 (0.64) & .056 \\
Condition: Minority $\times$ Task: Normative & 2.57 (0.91) & .005** & 0.10 (0.65) & .879 & 3.25 (0.79) & $<.001^{***}$ & 1.53 (0.63) & .016* \\
Condition: Majority $\times$ Time: $T_1$ & 0.28 (1.59) & .860 & -1.10 (1.14) & .334 & -0.90 (1.39) & .516 & -0.83 (1.11) & .455 \\
Condition: Minority $\times$ Time: $T_1$ & -0.68 (1.57) & .665 & 0.86 (1.13) & .447 & 1.19 (1.37) & .387 & 0.99 (1.10) & .370 \\
Condition: Majority $\times$ Time: $T_2$ & -0.26 (1.59) & .872 & -0.18 (1.14) & .875 & -0.59 (1.39) & .672 & 1.03 (1.11) & .355 \\
Condition: Minority $\times$ Time: $T_2$ & -0.93 (1.57) & .554 & 0.37 (1.13) & .744 & 0.09 (1.37) & .946 & -0.39 (1.10) & .721 \\
Condition: Majority $\times$ Time: $T_3$ & 0.12 (1.59) & .939 & 0.49 (1.14) & .666 & 0.64 (1.39) & .646 & 0.13 (1.11) & .906 \\
Condition: Minority $\times$ Time: $T_3$ & -0.29 (1.57) & .855 & -0.29 (1.13) & .798 & 0.15 (1.37) & .913 & -0.10 (1.10) & .927 \\
Task: Normative $\times$ Time: $T_1$ & -1.17 (1.11) & .295 & 1.53 (0.80) & .055 & 1.86 (0.97) & .056 & 1.84 (0.78) & .019* \\
Task: Normative $\times$ Time: $T_2$ & -0.45 (1.11) & .688 & 0.00 (0.80) & .999 & 0.14 (0.97) & .885 & 0.05 (0.78) & .953 \\
Task: Normative $\times$ Time: $T_3$ & 0.53 (1.11) & .635 & -0.64 (0.80) & .426 & -0.53 (0.97) & .582 & -1.06 (0.78) & .175 \\
Cond.~Majority $\times$ Task: Normative $\times$ Time: $T_1$ & -0.21 (1.59) & .896 & 2.07 (1.14) & .070 & 0.83 (1.39) & .548 & 1.46 (1.11) & .191 \\
Cond.~Minority $\times$ Task: Normative $\times$ Time: $T_1$ & -0.05 (1.57) & .974 & -1.66 (1.13) & .141 & -0.99 (1.37) & .470 & -1.55 (1.10) & .158 \\
Cond.~Majority $\times$ Task: Normative $\times$ Time: $T_2$ & -0.53 (1.59) & .741 & -0.24 (1.14) & .832 & -1.41 (1.39) & .309 & -1.35 (1.11) & .225 \\
Cond.~Minority $\times$ Task: Normative $\times$ Time: $T_2$ & 0.79 (1.57) & .616 & 0.04 (1.13) & .973 & 0.93 (1.37) & .498 & 0.62 (1.10) & .571 \\
Cond.~Majority $\times$ Task: Normative $\times$ Time: $T_3$ & 0.18 (1.59) & .909 & -0.92 (1.14) & .419 & -0.33 (1.39) & .811 & -1.01 (1.11) & .362 \\
Cond.~Minority $\times$ Task: Normative $\times$ Time: $T_3$ & 0.11 (1.57) & .942 & 0.67 (1.13) & .549 & -0.49 (1.37) & .721 & 0.31 (1.10) & .778 \\
\midrule
\multicolumn{9}{l}{\textit{Model fit and random effects}} \\
$N_{\text{obs}}$, $N_{\text{participants}}$ & \multicolumn{2}{c}{1016,\ 127} & \multicolumn{2}{c}{1016,\ 127} & \multicolumn{2}{c}{1016,\ 127} & \multicolumn{2}{c}{1016,\ 127} \\
Random intercept SD (participant) & \multicolumn{2}{c}{14.44} & \multicolumn{2}{c}{10.01} & \multicolumn{2}{c}{13.39} & \multicolumn{2}{c}{10.32} \\
Residual SD & \multicolumn{2}{c}{20.49} & \multicolumn{2}{c}{14.70} & \multicolumn{2}{c}{17.89} & \multicolumn{2}{c}{14.35} \\
AIC / BIC & \multicolumn{2}{c}{9281.7 / 9424.5} & \multicolumn{2}{c}{8599.2 / 8742.0} & \multicolumn{2}{c}{9017.8 / 9160.6} & \multicolumn{2}{c}{8562.0 / 8704.8} \\
\bottomrule
\end{tabular}
\label{tab:lmm_opinion_confidence_full}
\end{table*}

\begin{table}[htbp]
\centering
\caption{Distribution of opinion-shift trajectories by condition and task type. The table reports the number of participants classified as reinforcing, aligning, or oscillating within each social influence condition and task type, based on longitudinal opinion changes after baseline sign-flip preprocessing.}
\Description{The table summarizes how participants were distributed across three opinion-shift trajectories—reinforcing, aligning, and oscillating—by social influence condition and task type. Rows list Majority, Minority, and Diffusion conditions, each split into informative and normative tasks. The cells report the number of participants in each trajectory category, showing that all three conditions include participants who reinforced their initial views, aligned with AI positions, or oscillated over time, with oscillating trajectories generally being the most frequent across tasks.}
\small
\begin{tabular}{llrrr}
\toprule
Condition & Task Type & Reinforcing & Aligning & Oscillating \\
\midrule
Majority & Informative & 7 & 10 & 24 \\
 & Normative & 7 & 6 & 28 \\
Minority & Informative & 6 & 14 & 23 \\
 & Normative & 18 & 6 & 19 \\
Diffusion & Informative & 8 & 8 & 27 \\
 & Normative & 18 & 3 & 22 \\
\bottomrule
\end{tabular}

\label{tab:trajectories}
\end{table}

\clearpage
\section{Supplementary Figures}
\begin{figure*}[htbp]
  \centering
  \includegraphics[width=1.0\textwidth]{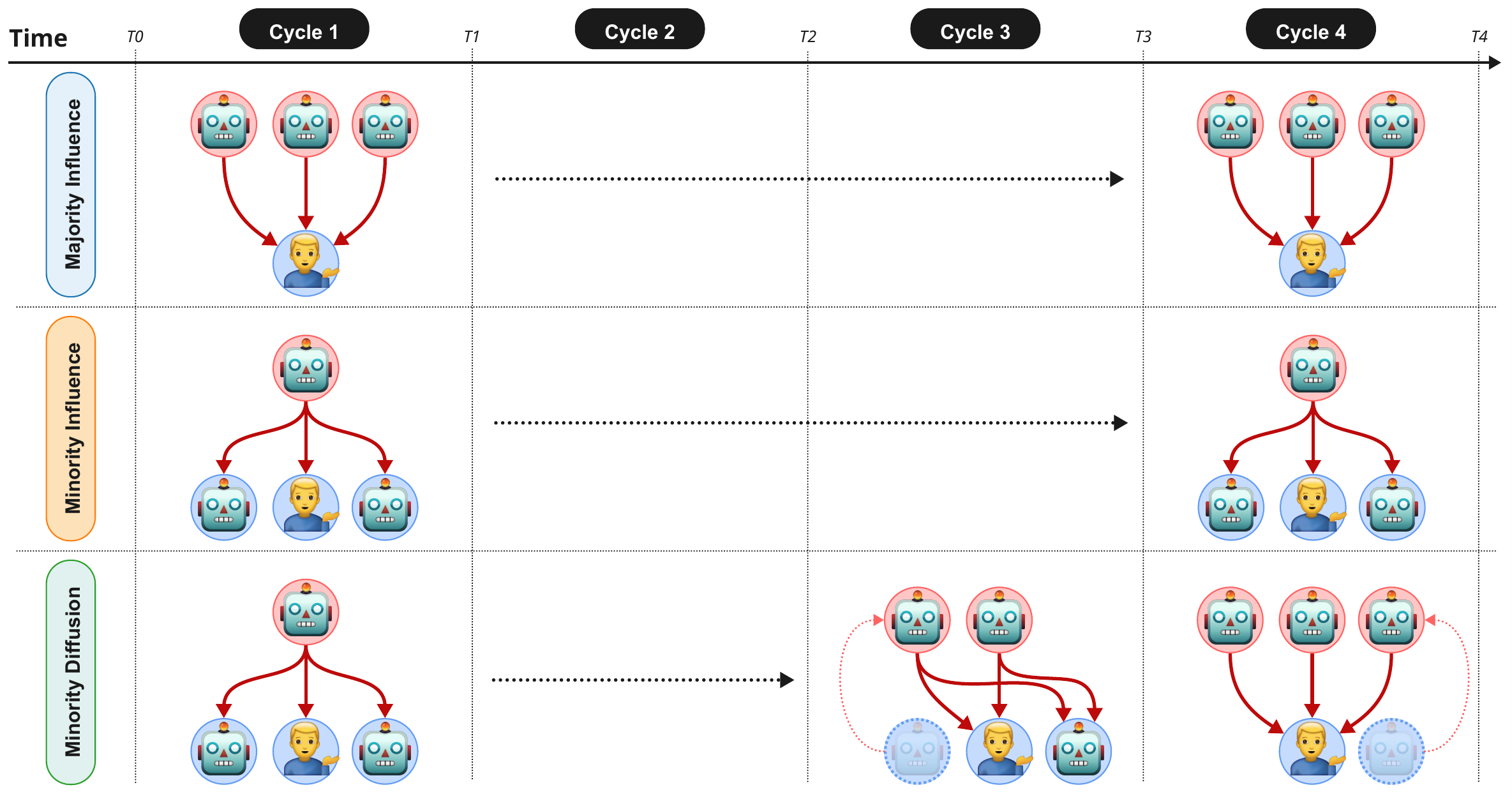}
  \caption{Timeline of the three experimental conditions. In the Majority Influence condition, all agents consistently opposed the participant across all cycles. In the Minority Influence condition, one dissenting agent opposed while two supported the participant throughout. In the Minority Diffusion condition, the session began with one minority agent, and additional agents gradually switched sides in later cycles, creating a new majority.}
  \Description{A timeline diagram compares three experimental conditions across four cycles (T0–T4). Majority Influence: Three red robots oppose the participant at every cycle. Minority Influence: One red robot dissents while two blue robots support the participant consistently across all cycles. Minority Diffusion: Starts with one red dissenting robot and two blue supportive robots. At cycle 3, one blue robot switches to dissent, and by cycle 4, both have switched, forming a new red majority.}
  \label{fig:experimentalDesign}
\end{figure*}

\begin{figure}[htbp]
  \centering
  \includegraphics[width=0.45\textwidth]{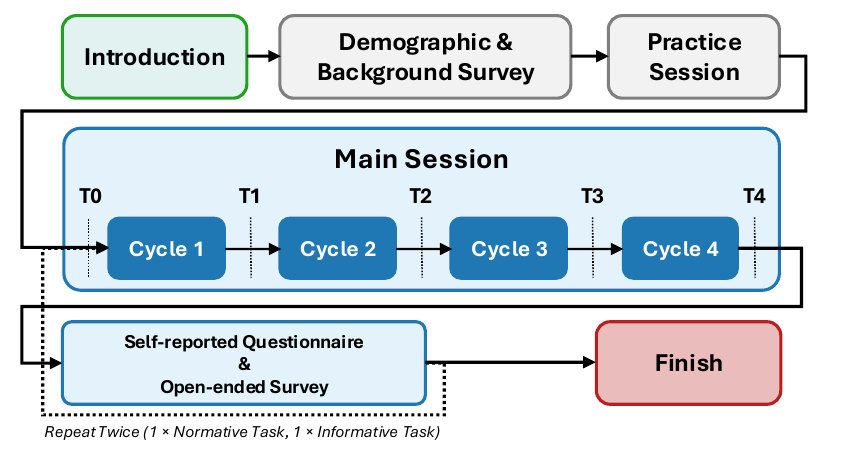}
  \caption{Experimental procedure. Participants completed an introduction, a background survey, and a practice session, followed by two main sessions (one normative and one informative task, each with four cycles). Afterward, they completed self-reported and open-ended surveys before concluding the study.}
  \Description{The diagram shows the experimental procedure as a flowchart. The sequence begins with the Introduction, followed by the Demographic and Background Survey, then the Practice Session. The Main Session consists of four repeated cycles, and participants complete this task twice, once with a normative task and once with an informative task. Afterward, they fill out a Self-reported Questionnaire and Open-ended Survey, and finally reach the Finish stage.}
  \label{fig:experimentalProcedure}
\end{figure}

\begin{figure*}[htbp]
  \centering
  \includegraphics[width=0.7\textwidth]{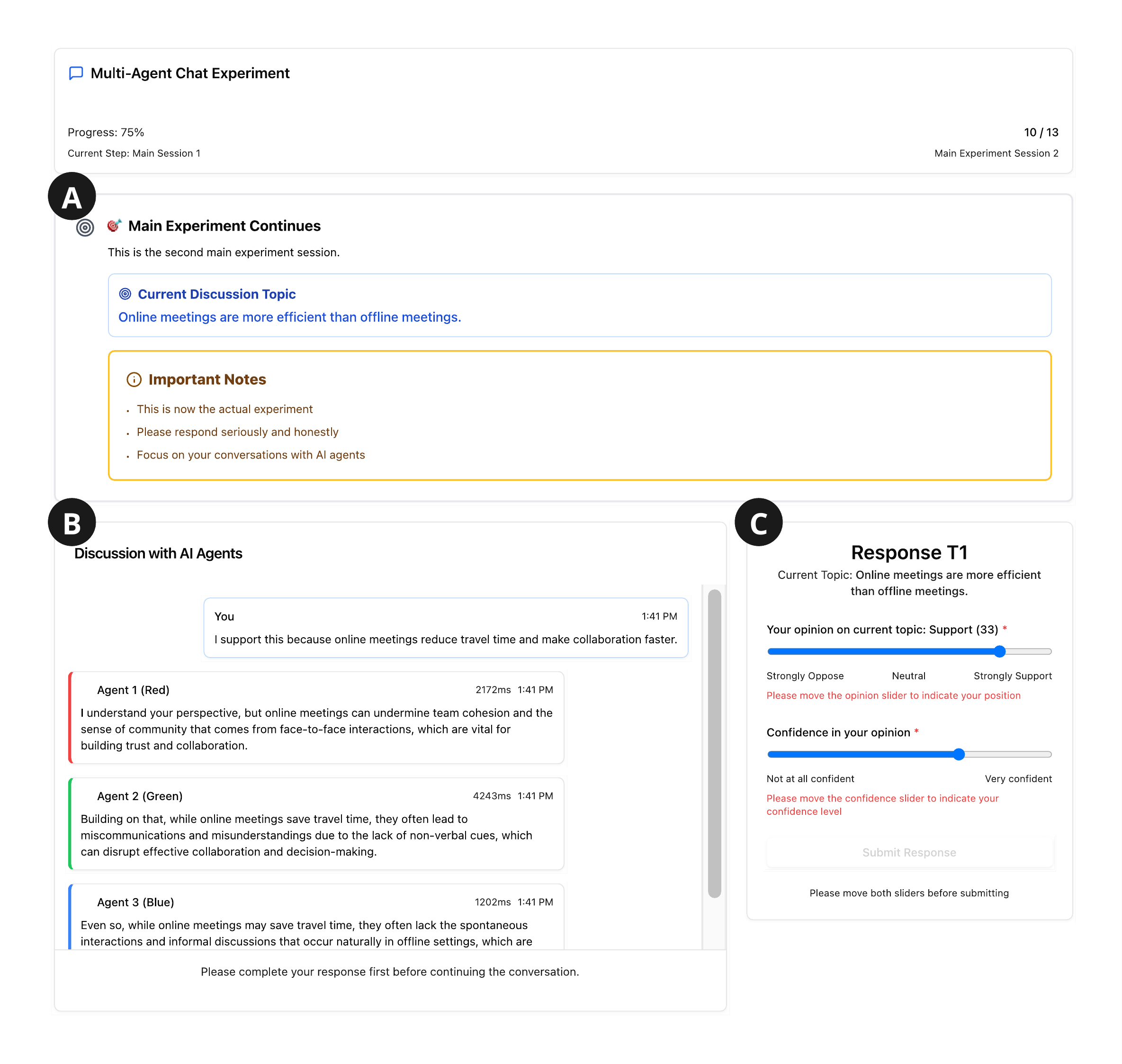}
  \caption{Experimental system interface. Panel A provides instructions and displays the current discussion topic. Panel B shows the group chat interface where participants interact with three AI agents. Panel C contains sliders for recording participants’ opinions and confidence as an in situ decision measure after each cycle.}
  \Description{The figure shows the user interface of the experimental system. At the top (Panel A), instructions are displayed with the current discussion topic “Online meetings are more efficient than offline meetings” and important notes reminding participants to respond seriously and focus on the AI conversation. In the middle (Panel B), a chat window shows a participant’s message supporting online meetings, followed by three AI agents responding with different arguments against this stance. At the right (Panel C), a panel titled “Response T1” includes two sliders: one for indicating the participant’s opinion from strongly oppose to strongly support, and another for rating confidence from not at all confident to very confident. Participants must move both sliders before submitting their response.}
  \label{fig:systemInterface}
\end{figure*}

\begin{figure*}[htbp]
  \centering
  \includegraphics[width=1.0\textwidth]{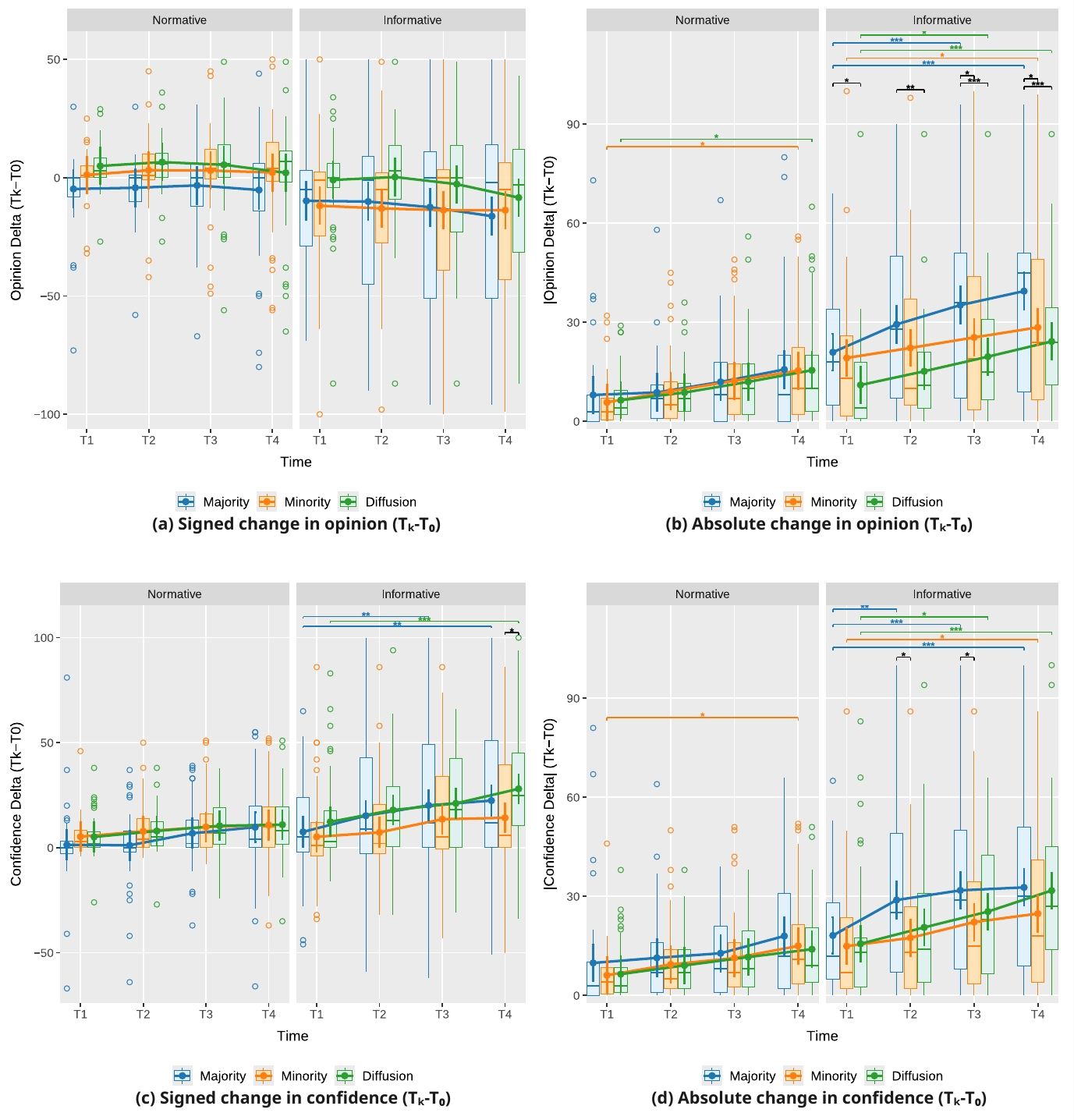}
  \caption{Opinion and confidence changes over time across majority, minority, and diffusion conditions, shown separately for normative and informative tasks. Panels (a–d) display signed and absolute deltas relative to $T_0$. Boxplots represent raw data, and overlaid lines show Estimated Marginal Means (EMMs) with 95\% confidence intervals (CIs). Bonferroni-significant contrasts are marked with brackets ($p < .05$\,*, $p < .01$\,**, $p < .001$\,***).}
  \Description{The figure shows four sets of boxplots with overlaid trend lines, illustrating how participants’ opinion and confidence changed over time relative to a baseline. Each row represents a different measure—signed opinion change, absolute opinion change, signed confidence change, and absolute confidence change—and each is displayed separately for normative tasks on the left and informative tasks on the right. Within every panel, three colors represent the three AI-agent conditions: Majority (blue), Minority (orange), and Diffusion (green). Boxplots show the distribution of individual participant changes at each time point, while smooth lines indicate estimated marginal means with confidence intervals. Across panels, informative tasks generally show larger shifts than normative tasks, and Minority and Diffusion conditions often display greater variability and steeper increases. Horizontal brackets mark statistically significant differences based on Bonferroni-corrected contrasts.}
  \label{fig:opinionConfidenceDelta_overall}
\end{figure*}